\documentstyle[emulateapj]{article}

\lefthead{Armitage \& Natarajan}
\righthead{Blandford-Znajek emission in low luminosity black hole systems}

\begin{document}

\submitted{ApJL in press}

\title{The Blandford-Znajek mechanism and emission from \\
       isolated accreting black holes}

\author{Philip J. Armitage and Priyamvada Natarajan}
\affil{Canadian Institute for Theoretical Astrophysics, McLennan Labs,
	60 St George St, Toronto, M5S 3H8, Canada}

\begin{abstract}
In the presence of a magnetic field, rotational energy can be
extracted from black holes via the Blandford-Znajek mechanism. We
use self-similar advection dominated accretion (ADAF) models to
estimate the efficiency of this mechanism for black holes accreting
from geometrically thick disks, in the light of recent
magnetohydrodynamic disk simulations, and show that the power from
electromagnetic energy extraction exceeds the accretion luminosity for
ADAFs at sufficiently low accretion rates. We
consider the detectability of isolated stellar
mass black holes accreting from the ISM, and show that for any 
rapidly rotating holes the efficiency of energy
extraction, $\epsilon_{\rm BZ} \equiv L_{\rm BZ} / \dot{M} c^2$, could
reach $10^{-2}$. The estimated total luminosity would be
consistent with the tentative identification of some EGRET sources as
accreting isolated black holes, if that energy is radiated primarily
as gamma rays. We discuss the 
importance of emission from the Blandford-Znajek mechanism for the
spectra of other advection dominated accretion flows, especially 
those in low luminosity galactic nuclei.
\end{abstract}	


\section{INTRODUCTION}

The formation of both stellar mass and supermassive black holes could well lead 
to rapidly rotating Kerr black holes. If so, this reservoir 
of rotational energy constitutes a potentially important energy 
source for driving jets and emission from black hole systems. 
Moreover, this energy can be extracted 
if the black hole is embedded in a magnetic field (Blandford \& Znajek 1977). 
Since all black holes of current astrophysical interest are probably accreting 
from magnetized disks, this has led to suggestions that 
the Blandford-Znajek process plays a vital role in
Active Galactic Nuclei (AGN) and other accreting black 
hole systems (Begelman, Blandford \& Rees 1984; Wilson \& Colbert 1995; 
Moderski, Sikora \& Lasota 1998; Paczynski 1998; Lee, Wijers \& Brown 1999). 

Recent papers (Ghosh \& Abramowicz 1997; Livio, Ogilvie, \& Pringle
1999; Li 1999) have re-evaluated the importance of the Blandford-Znajek
mechanism for AGN in light of advances in the understanding of
accretion disks -- in particular, the demonstration that
turbulence driven by magnetohydrodynamic instabilities (Balbus \&
Hawley 1991) leads to disk angular momentum
transport (Hawley, Gammie \& Balbus 1995; Brandenburg et al. 1995; 
Stone et al. 1996; for
a review see e.g. Hawley \& Balbus 1999). Two principal effects lead
to substantial downward revisions in the estimated efficiency of the
Blandford-Znajek mechanism for thin disks. First, in an accretion disk 
both the viscosity and the magnetic diffusivity are generated by
the same turbulent processes, and so have comparable magnitudes.
This implies that any external magnetic field can diffuse outward
through the inflowing gas, and will not accumulate on the hole.
Second, the strongest magnetic fields generated by a thin disk are
expected to have characteristic scales of order the disk scale height
$H$, with only weaker large scale fields threading the black hole and
contributing to the extraction of the rotational energy. 
Both arguments are weakened for thick disks for which $H \sim R$.

In this Letter, we consider the Blandford-Znajek contribution
to the luminosity of black hole systems at extremely low 
accretion rates. Motivated by the hard high energy spectra and 
low luminosity of such systems, Narayan \& Yi (1994; 1995) have 
developed an advection dominated accretion flow (ADAF) model, in 
which radiatively inefficient accretion occurs in a hot, geometrically thick 
geometry (see also Rees et al. 1982; 
Ichimaru 1977). Strong disk winds may also contribute to the observed 
dearth of emission (Blandford \& Begelman 1999).
As Rees et al. (1982) noted, the low 
radiative efficiency $\epsilon_{\rm ADAF} = L_{\rm ADAF} / \dot{M} c^2$ 
of ADAFs means that even rather weak emission arising from the 
Blandford-Znajek mechanism could make an important contribution to the 
overall luminosity. Here, we concentrate on the extreme case of 
isolated black holes accreting from the interstellar medium (ISM), 
and show that the power from the Blandford-Znajek process substantially 
improves the prospects for detecting any rapidly rotating holes.
We also comment on the implications 
for other very low luminosity accreting black holes, particularly 
those in quiescent galactic nuclei.

\section{EVALUATING THE BLANDFORD-ZNAJEK POWER}

For a black hole of mass $M$ and angular momentum $J$, with a magnetic
field $B_\bot$ normal to the horizon at $R_h$, the power arising from
the Blandford-Znajek mechanism is given by (e.g. Ghosh \& Abramowicz 1997; 
Macdonald \& Thorne 1982; Thorne, Price \& Macdonald 1986),
\begin{equation}
 L_{\rm BZ} = {1 \over 32} \omega_F^2 B_\bot^2 R_h^2 c 
 \left( {J \over J_{\rm max}} \right)^2
\label{eq1} 
\end{equation}
where $J_{\rm max} = GM^2 / c$ is the maximal angular momentum of the
hole. The factor $\omega_F^2 \equiv \Omega_F (\Omega_h - \Omega_F) /
\Omega_h^2$ depends on the angular velocity of
field lines $\Omega_F$ relative to that of the hole,
$\Omega_h$. Conventionally, we will assume that $\omega_F = 1/2$,
which maximizes the power output (Macdonald \& Thorne 1982; Phinney
1983; Thorne, Price \& Macdonald 1986).

To estimate $L_{\rm BZ}$ we need to relate 
$B_\bot$ to the conditions at the inner edge of an ADAF. 
First though, we briefly consider the possibility that $B_\bot$ 
at the black hole horizon could 
exceed the disk field at the inner disk edge, if
the disk is able to advect an organized external magnetic
field inwards. Whether this is possible depends on the relative timescales for
inward advection, controlled by the disk shear viscosity $\nu$, versus
outward diffusion, controlled by the magnetic diffusivity $\eta$.  For
a disk with magnetic Prantl number ${\cal{P}} \equiv \nu / \eta$, the
ability of the disk to advect field lines inwards depends on the
dimensionless parameter ${\cal{D}} = ( R / H ) {\cal{P}}$, where 
$H$ is the disk scale height at radius $R$. Lubow, Papaloizou \& 
Pringle (1994) found that ${\cal{D}} \lesssim 1$ was a necessary 
condition for strong inward field dragging (see also Reyes-Ruiz \& 
Stepinski 1996). Since it is generally believed that in a turbulent 
medium ${\cal{P}} \simeq 1$ (e.g. Parker 1979), this result implies that 
for a thin disk ${\cal{D}} \sim (R/H) \gg 1$, and external fields 
cannot accumulate at small radii. For thick disks, where $H \sim R$, 
the situation is less clear, but for a lower limit on $B_\bot$ 
we will assume that even in thick disks, advection of the field is ineffective. 
The field at the black hole then reflects only the local disk magnetic 
field strength. 
 
To estimate the physical conditions in the inner part of the disk, 
we employ the vertically averaged, self-similar ADAF solution of 
Narayan \& Yi (1994; Spruit et al. 1987). This solution assumes a 
Shakura-Sunyaev (1973) prescription for the shear viscosity, 
$\nu = \alpha c_s H$, where $\alpha$ is a free parameter and 
$c_s$ is the local disk sound speed. The character of the 
solution depends on the parameter $\epsilon^\prime$, where
\begin{equation}
  \epsilon^\prime \equiv { {5/3 - \gamma} \over { f (\gamma - 1) } },
\end{equation} 
and $\gamma$ is the ratio of specific heats. The factor $f$ 
measures the efficiency of radiative cooling, so that $f=1$ corresponds 
to the limit of no cooling at all, while $f=0$ resembles the case 
of a thin disk in which cooling is efficient. For $\alpha \ll 1$, 
as suggested by numerical simulations, the solution for the sound speed 
and the radial velocity $v_R$ is (Narayan \& Yi 1994),
\begin{equation} 
  v_R    \simeq  - { {3 \alpha} \over {5 + 2 \epsilon^\prime} } v_k, \, \,
  c_s^2  \simeq  { 2 \over {5 + 2 \epsilon^\prime} } v_k^2,
\end{equation}  
where $v_k = (GM / R)^{1/2}$ is the Keplerian velocity. Making use 
of $H / R = c_s / v_k$, and the continuity equation, 
\begin{equation} 
  \rho = - \dot{M} / 4 \pi R H v_R,
\end{equation} 
we find that the pressure $P = \rho c_s^2$ is given by, 
\begin{equation} 
 P = { {\dot{M} \sqrt{2}} \over {12 \pi \alpha} } (5 + 2 \epsilon^\prime)^{1/2} 
 (GM)^{1/2} R^{-5/2}.
\end{equation}
Studies that have dropped the requirements of
self-similarity and vertical averaging have found that this 
solution provides a good approximation to
ADAF flows (Narayan \& Yi 1995).

Numerical simulations of thin, magnetized accretion disks (Hawley, Gammie \&
Balbus 1995; Brandenburg et al. 1995; Stone et al. 1996), have shown that 
angular momentum transport is dominated by Maxwell
stresses. The magnetic contribution to $\alpha$
typically exceeds the fluid stresses by an order of magnitude, so that 
$\alpha \simeq \alpha_{\rm magnetic}$. The dominant field
component is toroidal, with saturation occurring when $P_{\rm mag} \ll
P$. Initial results from global simulations (Armitage 1998) suggest
that Narayan \& Yi's assumption -- that in thick disks the 
magnetic fields possess most of their power in long wavelength
modes of scale $\sim R$ -- is reasonable. 

We assume that $P_{\rm mag} = B^2 / 8 \pi \sim \alpha P$ 
in the inner disk, and make the important assumption that $B_\bot \approx B$. 
We further assume that the energy extracted from the hole is {\em not} 
all dumped into the internal energy of the accreting gas and lost invisibly 
across the horizon -- the fate of the bulk of the accretion energy (Narayan, 
Garcia \& McClintock 1997). This is reasonable provided that some fraction of the field 
lines threading the hole are open, for example as a result of a wind or a jet.
Evaluating the Blandford-Znajek power for the case of a maximally rotating hole, 
where $J = J_{\rm max}$ and $R_h = GM / c^2$, we then obtain,
\begin{equation} 
 L_{\rm BZ} \simeq { \sqrt{2} \over 192 } (5 + 2 \epsilon^\prime)^{1/2} 
 \dot{M} c^2. 
\end{equation} 
This corresponds to a constant efficiency, $\epsilon_{\rm BZ} \equiv 
L_{\rm BZ} / \dot{M} c^2$, given for the range of thick disk models by,
\begin{eqnarray}
  \epsilon_{\rm BZ} & \simeq & { \sqrt{14} \over 192} \ \  {\rm for} \ \epsilon^\prime=1 \\
  \epsilon_{\rm BZ} & \simeq & { \sqrt{10} \over 192} \ \   {\rm for} \ \epsilon^\prime=0  
\end{eqnarray} 
This efficiency drops rapidly for more slowly spinning holes, due 
to both the $(J / J_{\rm max})^2$ factor in equation
(\ref{eq1}), and because the pressure in the disk, which should be
evaluated at the marginally stable orbit, drops as
$R^{-5/2}$. Nonetheless, it suggests that the Blandford-Znajek
mechanism will be an important power source in the overall
energy budget of ADAFs around rapidly rotating black holes, with
efficiencies of up to $\sim 10^{-2}$. We note that if $\epsilon^\prime > 1$ 
(corresponding to a thin disk), these scalings suggest still higher 
efficiencies. More realistically, the suppressing effects 
identified by Livio, Ogilvie \& Pringle (1999) 
are expected to be important in this regime. Of course, if 
a thin disk existed the holes would be readily detectable 
due to the high radiative efficiency ($\sim 0.1$) of thin disk 
accretion. Additional electromagnetic extraction of energy
from the inner {\em disk}, which Livio, Ogilvie \& Pringle (1999)
argue is likely to exceed that from the hole itself, would also 
boost the efficiency.
  
\section{ISOLATED BLACK HOLES}

The known neutron stars and stellar mass black holes in mass transfer
binaries must be greatly outnumbered by the isolated populations of
such remnants, accreting only at a very low rate
from the ISM (e.g. Blaes \& Madau 1993). To date, only a few isolated
neutron stars have been identified (Walter \& Matthews 1997; Neuhauser
\& Trumper 1999). Identification of the corresponding black hole
population would provide both a record of massive star
formation in the galaxy, and insight into the behavior of 
accretion flows at extremely low $\dot{M}$.

The accretion rate onto a black hole accreting from the ISM is given 
by the Bondi-Hoyle rate (Hoyle \& Lyttleton 1939; Bondi \& Hoyle 1944). 
For a hole moving with velocity $v_\infty$ through a uniform medium 
with sound speed $c_\infty$, the accretion radius is,
\begin{equation} 
 R_a = { {2GM} \over {v_\infty^2 + c_\infty^2} }.
\end{equation} 
Gas falling within a cylinder of this radius is accreted, so that the accretion 
rate is $\dot{M}_{\rm Bondi} = \pi R_a^2 \rho_\infty \sqrt{v_\infty^2 + c_\infty^2}$, 
where $\rho_\infty$ is the density far upstream of the accretor.

The most luminous black holes are expected to be those accreting from
nearby, dense, molecular clouds (Fujita et al. 1998). For  
gas at temperatures of $T \sim 10^2 \ {\rm K}$, $v_\infty \gg c_\infty$, 
and the accretion rate is,
\begin{eqnarray}
 \dot{M}_{\rm Bondi} = 7.3 \times 10^{15} 
 \left( {M \over {10 M_\odot}} \right)^2
 \left( { n_\infty \over {10^2 \ {\rm cm}^{-3}} } \right) \nonumber \\
 \left( { v_\infty \over {10 \ {\rm kms}^{-1} } } \right)^{-3} \ {\rm gs}^{-1}
\end{eqnarray}
where we have taken a density $n_\infty$ (in particles / cm$^3$)
typical for a molecular cloud. The accretion rate will be much lower for 
holes accreting from hotter phases of the ISM. For gas at $n_\infty \sim 1 \
{\rm cm}^{-3}$ and $T \sim 10^4 \ {\rm K}$, the same hole accreting 
transonically ($v_\infty \approx c_\infty \approx 15 \ {\rm kms}^{-1}$) will 
have an accretion rate, 
$\dot{M}_{\rm Bondi} = 8 \times 10^{12} \ {\rm gs}^{-1}$. 

Some black holes in mass transfer binaries are believed to be rotating 
rapidly (Zhang, Cui \& Chen 1997), and we assume that some fraction 
of isolated black holes also have rapid ($J \sim J_{\rm max}$) rotation. 
Taking the Blandford-Znajek efficiency calculated earlier for the $\epsilon^\prime=1$ 
ADAF model (appropriate for a fully advection dominated, $\gamma = 4/3$ 
gas), the estimated power is,
\begin{eqnarray}
  L_{\rm BZ} & \simeq & 1.3 \times 10^{35} \ {\rm ergs}^{-1} \nonumber \\
    (n_\infty &=& 10^2 \ {\rm cm}^{-3}, v_\infty = 10 \ {\rm kms}^{-1}) \nonumber \\
  L_{\rm BZ} & \simeq & 1.4 \times 10^{32} \ {\rm ergs}^{-1} \nonumber \\
    (n_\infty &=& 1 \ {\rm cm}^{-3}, v_\infty = c_\infty = 15 \ {\rm kms}^{-1}). 
\label{eq_lum}      
\end{eqnarray}
These luminosities are comparable to those predicted on the basis of
models of spherical accretion of magnetized gas (Heckler \& Kolb 1996;
Ipser \& Price 1982, 1983). Spherical models are unrealistic 
at least for black holes accreting at high Mach numbers.
The luminosities in equation (\ref{eq_lum}) are considerably higher 
than estimates based on the radiative efficiency of an ADAF (Fujita et al. 1998), 
which is very low for small $\dot{M}$ (Narayan \& Yi 1995; Mahadevan \&
Quataert 1997). We note that at these accretion rates, the spindown time for 
the hole, even including the Blandford-Znajek losses, remains much greater than 
a Hubble time (e.g. King \& Kolb 1999).

Our estimates suggest that the prospects for detecting isolated
black holes are substantially better than would be inferred on the
basis of the low radiative efficiency of an ADAF, {\em if} a
reasonable fraction of the holes are rapidly rotating. Indeed,
these estimates are {\em energetically} consistent with associating isolated
black holes with some unidentified EGRET gamma ray sources (Lamb \& Macomb 1997; 
Dermer 1997; Yadigaroglu \& Romani 1997). Dermer (1997) estimates that these
sources have characteristic luminosities, in the 100 MeV - 5 GeV band,
of $\sim 2.5 \times 10^{35} \ {\rm ergs}^{-1}$ (for low latitude
sources with $\vert b \vert < 10^\circ$), and $\sim 6 \times 10^{32} \
{\rm ergs}^{-1}$ (for sources with $\vert b \vert > 10^\circ$). This
would be roughly consistent with the estimates for rotating $10
M_\odot$ black holes, provided that the efficiency of gamma ray 
emission was very high. High energy emission in AGN 
occurs via similar processes (e.g. Levinson \& Blandford 1996).  
We note that the rough 
proportionality of the luminosity of the unidentified sources on the
ISM density (Dermer 1997) would be consistent with the approximately
constant efficiency of the Blandford-Znajek mechanism.  It would {\em
not} be consistent with an ADAF model, whose radiative efficiency
declines at low $\dot{M}$ (Narayan \& Yi 1995).

\section{ADVECTION DOMINATED FLOWS IN GALACTIC NUCLEI}

Similar considerations apply to other low accretion rate flows that
are advection dominated. The radiative efficiency of an
ADAF is given approximately by,
\begin{equation} 
 \epsilon_{\rm ADAF} = 0.1 \left( {\dot{m} \over \alpha^2} \right),
\end{equation}
where $\dot{m} \equiv \dot{M} / \dot{M}_{\rm Edd}$, and the Eddington
accretion rate for an electron scattering opacity $\kappa_{\rm es}$,
and fiducial efficiency $\eta_{\rm Eff}=0.1$, is defined as
$\dot{M}_{\rm Edd} = 4 \pi GM / \eta_{\rm Eff} \kappa_{\em es} c$ 
(Narayan \& Yi 1995). For an estimated 
$\alpha = 0.1$, this implies that the Blandford-Znajek effect 
could be an important contributor to the overall power output for $\dot{m}
\lesssim 2 \times 10^{-3}$.

In addition to isolated black holes in the galactic disk, some
supermassive black holes in quiescent galactic nuclei also 
fall into this regime (Fabian \& Rees 1995). The weak emission from
the nuclei of these galaxies provides both a test of
the ADAF model (Di Matteo et al. 1999a), and a
possible contributor to the X-ray background (Di Matteo \& Fabian
1997; Di Matteo et al. 1999b).  Taking our estimates at face value, 
the good agreement of computed ADAF spectra with
observations in some systems -- especially the
galactic center (Mahadevan 1998) -- would be consistent with the black
holes in those systems {\em not} being close to maximally rotating.

\section{DISCUSSION}

In this Letter, we have discussed the importance of the 
Blandford-Znajek (1977) mechanism for ADAFs at low accretion rates. 
For these thick disks, which are similar 
to the tori originally considered by Rees et al. (1982), the arguments 
advanced for the likely irrelevance of the Blandford-Znajek mechanism 
(Ghosh \& Abramowicz 1997; Livio, Ogilvie \& Pringle 1999) are considerably 
weaker than for thin disks. Even with 
pessimistic assumptions, the total electromagnetic power 
liberated from the hole or inner disk will exceed the radiative 
luminosity for sufficiently low accretion rates. For maximally 
rotating black holes, and self-similar ADAF models, we obtain the criteria
$\dot{m} = \dot{M} / \dot{M}_{\rm Edd} \lesssim 2 \times 10^{-3}$. Additional 
Blandford-Znajek power
could then be important for the spectra of ADAFs, both for supermassive 
black holes in the apparently quiescent cores of elliptical galaxies
(Allen, Di Matteo \& Fabian 1999), 
and for stellar mass black holes that are similarly starved of fuel.

For the extreme case of isolated stellar mass black holes accreting from 
the local ISM, these results imply that the detectability of any
rapidly rotating holes is substantially better than estimates 
based on pure ADAF models (Fujita et al. 1998). The 
inferred luminosities are similar to those suggested for some unidentified 
EGRET sources (Dermer 1997), if we assume (very optimistically) that most of the 
power from the hole is ultimately radiated as gamma-ray emission. With 
better signal to noise observations, variability is likely to provide 
additional constraints on models for these EGRET sources. 
Isolated stellar mass black holes, like those in mass transfer binaries,
will display variability on all timescales exceeding the shortest dynamical 
timescales (of the order of milliseconds) available in the system. 
For holes accreting at high Mach numbers there may also be additional 
quasi-periodic variations arising from `flip-flop' type instabilities 
in Bondi-Hoyle accretion, though the persistence of 
these features in three dimensional geometry (Ruffert 1997) as opposed  
to axisymmetric calculations (Benensohn, Lamb \& Taam 1997), is 
not certain. These instabilities have characteristic 
timescales of $\sim R_a / c_\infty$, the sound crossing time at the 
Bondi accretion radius. These variability signatures 
apply both to the high energy emission, and to emission in the 
optical wavebands that may be 
detectable via current wide area sky surveys (Heckler \& Kolb 1996).

\acknowledgements

We thank Brad Hansen, Norm Murray, Gordon Ogilvie and the referee for useful 
discussions on these topics.


\begin{references}

Allen, S. W., Di Matteo, T., \& Fabian, A. C., 1999, MNRAS, submitted, 
astro-ph/9905053 

Armitage, P. J., 1998, ApJ, 501, L189

Balbus, S. A., \& Hawley, J. F. 1991, ApJ, 376, 214

Begelman, M. C., Blandford, R. D., \& Rees, M. J., 1984, Reviews of
Mod. Phys., Vol. 56, Part I, p.~255

Benensohn, J. S., Lamb, D. Q., \& Taam, R. E., 1997, ApJ, 478, 723

Blaes, O. \& Madau, P., 1993, ApJ, 403, 690

Blandford, R. D., \& Begelman, M. C., 1999, MNRAS, 303, L1

Blandford, R. D., \& Znajek, R. L., 1977, MNRAS, 179, 433

Bondi, H., \& Hoyle, F., 1944, MNRAS, 104, 273

Brandenburg, A., Nordlund, A., Stein, R. F., \& Torkelsson, U. 1995,
ApJ, 446, 741

Dermer, C. D., 1997, Proc. of the Fourth Compton Symp., eds. Dermer, Strickman
\& Kurfess, AIP Conf. Proc. 410, p. 1275. 

Di Matteo, T., \& Fabian, A. C., 1997, MNRAS, 286, 393

Di Matteo, T., Esin, A., Fabian, A. C., \& Narayan, R., 1999b, MNRAS, 305, L1

Di Matteo, T., Fabian, A. C., Rees, M. J., Carilli, C. L., \& Ivison, R. J., 
1999a, MNRAS, 305, 492

Fabian, A. C., \& Rees, M. J., 1995, MNRAS, 277, L55

Fujita, Y., Inoue, S., Manmoto, T., \& Nakamura, K. E., 1998, ApJ, 495, L85 

Ghosh, P., \& Abramowicz, M., 1997, MNRAS, 292, 887

Hawley, J. F., \& Balbus, S. A., 1999, Astrophysical Discs, eds. Sellwood
\& Goodman, ASP Conf. Series, 160, p.~108  

Hawley, J. F., Gammie, C. F., \& Balbus, S. A. 1995, ApJ, 440, 742

Heckler, A. F., \& Kolb, E. W., 1996, ApJ, 472, L85 

Hoyle, F., \& Lyttleton, R. A., 1939, Proc. Cambridge Philos. Soc., 
35, 405 

Ichimaru, S., 1977, ApJ, 214, 840

Ipser, J. R., \& Price, R. H., 1982, ApJ, 255, 654

Ipser, J. R., \& Price, R. H., 1983, ApJ, 267, 371

King, A. R., \& Kolb, U., 1999, MNRAS, 305, 654

Lamb, R. C., \& Macomb, D. J., 1997, ApJ, 488, 872

Lee, H. K., Wijers, R. A. M. J., \& Brown, G., 1999, Physics Reports, 
submitted, astro-ph/9906213

Levinson, A., \& Blandford, R. D., 1996, ApJ, 456, L29

Li, L.-X., 1999, ApJL, submitted, astro-ph/9902352 

Livio, M., Ogilvie, G., \& Pringle, J. E., 1999, ApJ, 512, 100

Lubow, S. A., Papaloizou, J., \& Pringle, J. E., 1994, MNRAS, 267, 235

Macdonald, D. D., \& Thorne, K. S., 1982, MNRAS, 198, 345

Mahadevan, R., \& Quataert, E., 1997, ApJ, 490, 605

Mahadevan, R., 1998, Nature, 394, 651

Moderski, R., \& Sikora, M., \& Lasota, J.-P., 1998, MNRAS, 301, 142

Narayan, R., Garcia, M. R., \& McClintock, J. E., 1997, ApJ, 478, L79

Narayan, R., \& Yi, I., 1994, ApJ, 428, L13

Narayan, R., \& Yi, I., 1995, ApJ, 452, 710

Neuhauser, R., \& Trumper, J. E., 1999, A\&A, 343, 151

Paczynski, B., 1998, ApJ, 494, L45

Parker, E. N., 1979, Cosmical Magnetic Fields. Clarendon Press, Oxford

Phinney, E. S., 1983, Ph.D. Thesis, University of Cambridge

Rees, M. J., Phinney, E. S., Begelman, M. C., \& Blandford, R. D., 1982,
Nature, 295, 17

Reyes-Ruiz, M., \& Stepinski, T. F., 1996, ApJ, 459, 653

Ruffert, M., 1997, A\&A, 317, 793

Shakura, N. I., \& Sunyaev, R. A. 1973, A\&A, 24, 337

Spruit, H. C., Matsuda, T., Inoue, M., \& Sawada, K., 
1987, MNRAS, 229, 517 

Stone, J. M., Hawley, J. F., Gammie, C. F., \& Balbus, S. A., 1996, ApJ, 
463, 656

Thorne, K. S., Price, R. H., \& Macdonald, D. D., 1986, Black Holes: The
Membrane Paradigm. Yale Univ. Press, New Haven CN, p. 132

Walter, F. M., \& Mathews, L. D., 1997, Nature, 389, 358

Wilson, A. S., \& Colbert, E. J. M., 1995, ApJ, 438, 62

Yadigaroglu, I.-A., \& Romani, R. W., 1997, ApJ, 476, 347 

Zhang, S. N., Cui, W., \& Chen W., 1997, ApJ, 482, L155

\end{references}
\end{document}